\documentclass[conference]{IEEEtran}
\IEEEoverridecommandlockouts
\usepackage{cite}
\usepackage{amsmath,amssymb,amsfonts}
\usepackage{algorithmic}
\usepackage{graphicx}
\usepackage{textcomp}
\usepackage{xcolor}
\def\BibTeX{{\rm B\kern-.05em{\sc i\kern-.025em b}\kern-.08em
    T\kern-.1667em\lower.7ex\hbox{E}\kern-.125emX}}
\begin{document}

\IEEEoverridecommandlockouts
\IEEEpubid{\begin{minipage}[t]{\textwidth}\ \\[10pt]
        \centering\normalsize{979-8-3315-8170-1/26/\$31.00 \copyright 2026 IEEE}
        \hspace{\columnsep}\makebox[\columnwidth]{\parbox{\columnwidth}{\vspace{6pt}\scriptsize \copyright 2026 IEEE. Personal use of this material is permitted. Permission from IEEE must be obtained for all other uses, in any current or future media, including reprinting/republishing this material for advertising or promotional purposes, creating new collective works, for resale or redistribution to servers or lists, or reuse of any copyrighted component of this work in other works. DOI: 10.1109/ICHQP67927.2026.11756967}}
\end{minipage}}

\title{Assessment of Errors of Fundamental Frequency Estimation Methods in the Presence of Voltage Fluctuations and Distortions 
\thanks{This research was funded in whole or in part by National Science Centre,
Poland – 2024/55/D/ST7/00441. For the purpose of Open Access, the author has applied a~CC BY public copyright licence to any Author Accepted Manuscript (AAM) version arising from this submission.}
}

\makeatletter
\newcommand{\linebreakand}{%
  \end{@IEEEauthorhalign}
  \hfill\mbox{}\par
  \mbox{}\hfill\begin{@IEEEauthorhalign}
}
\makeatother

\author{
\IEEEauthorblockN{Antonio Bracale, Pasquale De Falco}
\IEEEauthorblockA{\textit{Department of Engineering} \\
University of Naples Parthenope\\
Naples, Italy} 
\and
\IEEEauthorblockN{Piotr Kuwa{\l}ek, Grzegorz Wiczy{\' n}ski}
\IEEEauthorblockA{\textit{Institute of Electrical Engineering and Electronics} \\
Pozna{\' n} University of Technology\\
Pozna{\' n}, Poland \\
piotr.kuwalek@put.poznan.pl}
}

\maketitle

\begin{abstract}
The fundamental frequency is one of the parameters that define power quality. Correctly determining this parameter under the conditions that prevail in modern power grids is crucial. Diagnostic purposes often require an efficient estimation of this parameter within short time windows. Therefore, this article presents the results of numerical simulation studies that allow the assessment of errors in various fundamental frequency estimation methods, including the standard IEC 61000-4-30 method, when the analyzed signal has a form similar to that found in modern power grids. For the purposes of this study, a test signal was adopted recreating the states of the power grid, including the simultaneous occurrence of voltage fluctuations and distortions. Conclusions are presented based on conducted research.
\end{abstract}


\section{Introduction}

Power quality disturbances are a common phenomenon in modern power grids. According to the current benchmark report on power quality~\cite{PQRep}, for which the level of power quality disturbances was globally monitored, one of the most common phenomena is voltage fluctuations and distortions. This condition results in the analyzed voltage signal often being complex and nonstationary. For such a signal, it is necessary to properly determine the power quality parameters, including the fundamental frequency~\cite{f0_1,f0_2,f0_3}. According to current regulatory requirements, according to the standard EN 50160, the fundamental frequency value should be between 47 and 52 Hz for 100\% of the week and between 49.5 Hz and 50.5 Hz for 99.5\% of the year. It is worth noting that this standard specifies that the fundamental frequency value should be determined over a 10-second interval to verify the measured value's compliance with permissible values. 

\begin{figure}[htbp]
\centerline{\includegraphics[width=.85\columnwidth]{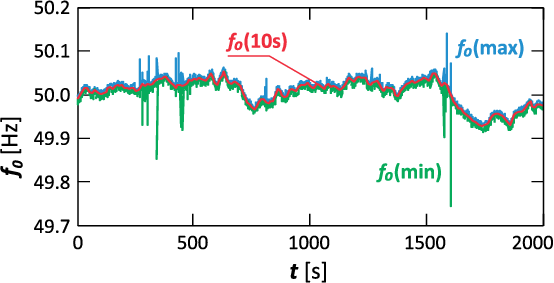}}
\caption{An example of the fundamental frequency $f_0$ over time characteristic in a power network recorded using PQA, where the red line shows the fundamental frequencies $f_0$(10s) measured at a 10-second interval, and the blue and green lines show the maximum $f_0$(max) and minimum $f_0$(min) fundamental frequencies measured at successive 200~ms intervals within a 10-second window. }
\label{fig1}
\end{figure}

Currently, power quality analyzers are mainly used to measure parameters that determine power quality. Class A compliance requires measuring these parameters with a specified inaccuracy in accordance with the IEC 61000-4-30 standard. For fundamental frequency measurement, this standard specifies a procedure for estimating this quantity, but also allows for the use of other methods for estimating the fundamental frequency. However, in both cases, the standard does not specify a complete measurement chain for this quantity, which can lead to problems in assessing the fundamental frequency in complex voltage signals. The fundamental frequency determined over a 10-second period exhibits little variability in a real power grid. Significant deviations in the fundamental frequency determined over a 10-second period are often noticeable during a blackout. However, if the fundamental frequency were determined for a shorter time interval, for example, 200~ms (10 fundamental voltage cycles with a rated fundamental frequency of 50~Hz), greater variability of the fundamental frequency could be observed, as shown in Fig.~\ref{fig1} and Fig.~\ref{fig2}, which present sample field measurement results using a Class A Power Quality Analyzer (PQA). This variability is related to the dynamics of load changes in the network and the response of the system automation to the resulting load changes. Analysis of the fundamental frequency variability determined for a 200-ms window is often used in power quality diagnostics when identifying the source of disturbances. Therefore, it is important that the method used for fundamental frequency estimation is characterized by small estimation errors, even for complex signal forms.

\begin{figure}[htbp]
\centerline{\includegraphics[width=.85\columnwidth]{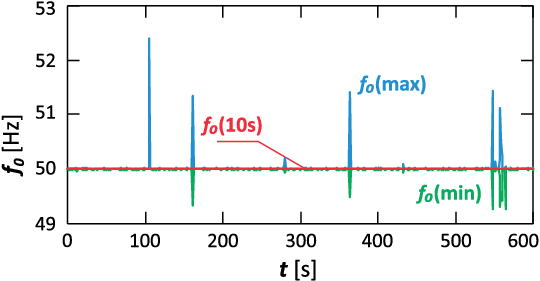}}
\caption{An example of the characteristic fundamental frequency $f_0$ over time on board a passenger train recorded using PQA, where the red line shows the fundamental frequencies $f_0$(10s) measured at a 10-second interval, and the blue and green lines show the maximum fundamental frequencies $f_0$(max) and minimum $f_0$(min) measured at successive 200~ms intervals within a 10-second window.}
\label{fig2}
\end{figure}

This paper presents assessment of fundamental frequency estimation errors for the method specified in the IEC standard and other estimation methods available in the literature. Numerical simulation studies were conducted using a signal recreating typical disturbances in a low-voltage power network. Section~\ref{sec2} describes the fundamental frequency estimation methods studied. Section~\ref{sec3} presents a description of the research. Section~\ref{sec4} presents selected research results and a discussion. Section~\ref{sec5} presents the final conclusions. 

\section{Considered methods of estimation of the fundamental frequency}\label{sec2}

\subsection{IEC 61000-4-30 Method}

The IEC 61000-4-30 standard specifies that the fundamental frequency should be determined according to the following relationship:
\begin{equation}
\widetilde{f_0} = \frac{N_{T_0}}{\Delta t_{N_{T_0}}},
\end{equation}
where $N_{T_0}$ is the total number of periods in the measurement window, $\Delta t_{N_{T_0}}$ is the duration of the total number of periods in the measurement window. Additionally, the standard specifies that, to minimize the impact of multiple zero crossings, harmonics and interharmonics should be suppressed before each estimation. Therefore, it can be concluded that, in the basic variant, the standard specifies that the start and end times of the periods should be determined using the zero-crossing detection method. It is also worth noting that, with respect to the aforementioned filter, there is no specific filter characteristic specified for this purpose. For the purposes of the study, a 100th-order band-pass filter with finite impulse response was adopted. The bandwidth was assumed to be 46-54~Hz.

\subsection{Autocorrelation method}

The autocorrelation method determines the fundamental frequency as the inverse of the fundamental period estimated by the method. In turn, the fundamental period using this method is determined as the average value of the difference in the local maxima of the autocorrelation function in the time domain, taking into account information about the sampling frequency. For a more stable operation of the method, when determining local maxima, it is assumed that the distance between local maxima cannot be less than approx. 13.3~ms.

\subsection{Hilbert method}

The Hilbert method determines the fundamental frequency for a given time interval as the average value of the instantaneous frequency $f_{\text{inst}}(t)$ in the analyzed window. The instantaneous frequency $f_{\text{inst}}(t)$ is determined according to the relationship: 

\begin{equation}
f_{\text{inst}}(t) = \frac{1}{2\pi} \frac{\text{d} \phi (t)}{\text{d} t},
\end{equation}
where $\phi (t)$ is the instantaneous phase determined as the argument of the analytical signal of the measured signal, the imaginary part of which is the Hilbert transform of the measured signal. 

\subsection{Modified ESPRIT method}

The ESPRIT algorithm models the sampled waveform as a superposition of complex exponentials. Specifically, given a sequence of sampled data $x$($n$) of length $L$, the signal is approximated as follows~\cite{AB1}:
\begin{equation}
\label{eq3}
\widehat{x}(n)=\sum_{k=1}^{M} A_k e^{j\psi_k } e^{(\alpha_k+j2\pi f_k\ )nT_s }+r(n),
\end{equation}
where $n$=0,1,…,$L$-1, $r$($n$) denotes additive white noise, and $A_k$, $\psi_k$, $f_k$, and $\alpha_k$ represent the amplitude, initial phase, frequency, and damping factor of the $k$-th component, respectively. These parameters are unknown and must be estimated. Defining $h_k=A_k e^{j\psi_k}$, (\ref{eq3}) can be rewritten in matrix form as:
\begin{equation}
\label{eq4}
\widehat{\mathbf{x}} = \mathbf{V \Phi ^ n H}  +\mathbf{r}, 
\end{equation}
where $\widehat{\mathbf{x}}$ and $\mathbf{r}$ are vectors containing $N$ consecutive samples ($N<L$), $\mathbf{H} =[h_1,\ldots,h_M]^T$, $\mathbf{V}$ is a Vandermonde matrix whose elements depend on the signal poles $e^{(\alpha_k+j2\pi f_k)T_s}$, $\mathbf{\Phi}= \text{diag} \left( e^{(\alpha_1+j2\pi f_1)T_s},\ldots,e^{(\alpha_M+j2\pi f_M)T_s}  \right)$ is referred to as the rotation matrix, and the number $N<L$ denotes the selected order of the correlation matrix. The ESPRIT method allows to find the solution of (\ref{eq4}) exploiting the shift-invariance property of the signal subspace.

Let $\widehat{\mathbf{S}}$ be the matrix of eigenvectors associated with the $M$ largest eigenvalues of the sample correlation matrix $\mathbf{R_x}$. Two submatrices $\widehat{\mathbf{S_1}}$ and $\widehat{\mathbf{S_2}}$ are formed by selecting the first and last $N$-1 rows of $\widehat{\mathbf{S}}$, respectively. The following relation holds: $\widehat{\mathbf{S_2}} = \widehat{\mathbf{S_1}} \mathbf{\Psi} $, where $\mathbf{\Psi}$ is a matrix that shares the same eigenvalues as $\mathbf{\Phi}$. An estimate of $\mathbf{\Psi}$ is obtained via the least-squares solution applying the following equation: 
\begin{equation}
\label{eq5}
\widehat{\Psi} = \left( \widehat{\mathbf{S_1}}^H\widehat{\mathbf{S_1}} \right)^{-1}\widehat{\mathbf{S_1}}^H\widehat{\mathbf{S_2}}. 
\end{equation}
where $H$ is the Hermitian transformation.

The damping factors and frequencies are estimated from the real and imaginary parts of the natural logarithm of eigenvalues $\widehat{\lambda_i}$, respectively. Once the frequencies and damping factors are known, the amplitudes and initial phases can be estimated using a least-squares approach~\cite{AB1}. The traditional sliding-window ESPRIT method provides accurate spectral estimates but suffers from high computational complexity, since the subspace decomposition, eigenvalue problem, and least-squares fitting must be solved for each analysis sliding window. To overcome this limitation, a modified scheme was proposed in~\cite{AB2}, aiming at preserving estimation accuracy while significantly reducing computational cost. This modified approach has been successfully applied to the assessment of both harmonic and supraharmonic waveform distortion in power systems in~\cite{AB3}.

\section{Research description}\label{sec3}

The research was conducted in the Matlab programming environment, where a test signal was numerically generated to recreate the actual supply voltage signal~\cite{utest} that can occur in a modern power system, and for which the fundamental frequency at specific times is known. This test signal can be described by the following relationship~\cite{utest}:
\begin{equation}
\label{eq7}
\begin{aligned}
&u_{\text{test}}(t)=[1+k_{\text{AM}}u_{\text{mod}}(t)]\cdot \\ &\cdot \sum_{h-1}^{H}U_{\text{amp}_h} \cos{\left( 2\pi h f_0 t+\phi_h +\frac{h\Delta f_0 2\pi }{|u_{\text{mod}}(t)|_{\text{max}}}\int{u_{\text{mod}}(t)\text{d}t}\right)} + \\ &+ \text{noise} , 
\end{aligned}
\end{equation}
where $k_{\text{AM}}$ is the amplitude modulation coefficient, $h$ is the harmonic order, $H$ is the maximum harmonic order considered, $f_0$ is the fundamental frequency of the carrier signal, $\phi_h$ is the initial phase of the $h$-th harmonic, $\Delta f_0$ is the deviation of the fundamental frequency, ${U_{\text{amp}}}_h$ is the amplitude of the $h$-th harmonic, noise is white noise with a specific SNR (Signal-to-Noise Ratio) expressed in decibels, $u_{\text{mod}}(t)$ is a modulating signal causing amplitude and frequency variation of a specific nature:
\begin{equation}
\label{eq8}
u_{\text{mod}}(t)=\text{sign}(\sin{(2\pi f_m t)}), 
\end{equation}
where $f_m$ is the modulating frequency. The presented signal is a model of simultaneously occurring signal fluctuations and distortions, for which it is possible to unambiguously determine the fundamental frequency. The harmonic values for the studies were assumed to be those for a ``clipped cosine" signal~\cite{Dist} for various clipped level $m_c$ ($m_c = 0.01, 0.8, 1$). The ``clipped cosine" signal~\cite{Dist} recreates typical distortions in low-voltage networks, which result from the interaction of the input stages of switched-mode power supplies. For $m_c$ = 1, the carrier signal is an ideal sinusoidal signal; for $m_c$ = 0.8, the signal is distorted with a THD value at the maximum allowed limit in a low-voltage network, and for $m_c$ = 0.01, the signal is significantly distorted (an extreme case, rarely encountered in a real rigid power grid, but typical for power supply in electric vehicles). The assumed modulating signal has step changes because in practice, load variability often occurs abruptly due to the sudden disconnection/connection of loads. The amplitude modulation coefficient $k_{\text{AM}}$ was assumed to be 0.05. The white noise introduced into the signal recreates nondeterministic load changes in the network. The study analyzes signals without noise and with noise at SNR levels of 10, 0, and -10~dB, respectively. The modulating frequencies $f_m$ adopted in the study, equal to 0.2, 0.5, 1, 2, 5, 10, and 20~Hz, allow assessing the effectiveness of the methods in the event of nonstationarity within the measurement window. Note that some values of $f_m$ lead to fundamental frequency values that are not observable in reality. However, use of these extreme values are useful for assessing the ability of the methods to track rapid frequency variations The base values of the fundamental frequency $f_0$ (carrier frequency) were assumed to be 47, 49.98, 50, 50.02, and 52~Hz, respectively. The frequency deviations $\Delta f_0$ were assumed to be 0.1, 1, and 10~Hz, respectively. 

The indicated test signal with the appropriate parameters was generated at a sampling rate $f_s$ of 327,680~Sa/s in a 10-second window, where the average fundamental frequency value for the 10-second window is close to the base value of the fundamental frequency~$f_0$. At the same time, within the indicated time window, the fundamental frequency changes in a manner determined by the modulating signal and the frequency deviation values. Therefore, in windows shorter than 10 seconds, the fundamental frequency significantly deviates from the base value of the fundamental frequency. For the purposes of the study, a 200~ms window was adopted for calculations, shifted approximately every 0.305~ms (100 samples). In each measurement window, the fundamental frequency was estimated using the methods specified in Section~\ref{sec2}, and the reference frequency ${f_0}_{\text{ref}}$ was determined as the average value from the 200~ms window for the known instantaneous fundamental frequency:
\begin{equation}
\label{eq9}
{f_0}_{inst}\left(t\right)=f_0\left(1+\Delta f_0u_{\text{mod}}\left(t\right)\right).
\end{equation}

For the purpose of assessing the errors of individual fundamental frequency estimation methods, the relative error given by the following relationship was assumed:
\begin{equation}
\label{eq10}
\delta f_0=\frac{\left|\widetilde{f_0}-{f_0}_{\text{ref}}\right|}{{f_0}_{\text{ref}}}, 
\end{equation}
where $\widetilde{f_0}$ is the value of the fundamental frequency estimated by the specific method under consideration in the specific 200~ms measurement window, and ${f_0}_{\text{ref}}$ is the reference value of the fundamental frequency.  

The specified constant sliding window of 200~ms applies to the IEC method and methods based on autocorrelation and the Hilbert transform. In the case of the modified ESPRIT method, due to the inherent characteristics of this method, the window duration is approximately 200~ms and is not always constant. For the modified ESPRIT method, the duration of the sliding window used for estimating the fundamental frequency is variable, and the window shift is also variable and linked to the adaptive features of this method. Therefore, it is worth emphasizing that, although the error can be defined in the same way (\ref{eq10}), the reference value of the fundamental frequency must be determined by taking into account the variable window duration.

\section{Research results and discussion}\label{sec4}

For the purpose of presenting the results, the following method designations are adopted: IEC 61000-4-30 Method applied with a 200~ms window as IEC, Autocorrelation method as xcorr, Hilbert method as Hilbert, Modified ESPRIT method as Esprit. 

\begin{figure}[htbp]
\centerline{\includegraphics[width=\columnwidth]{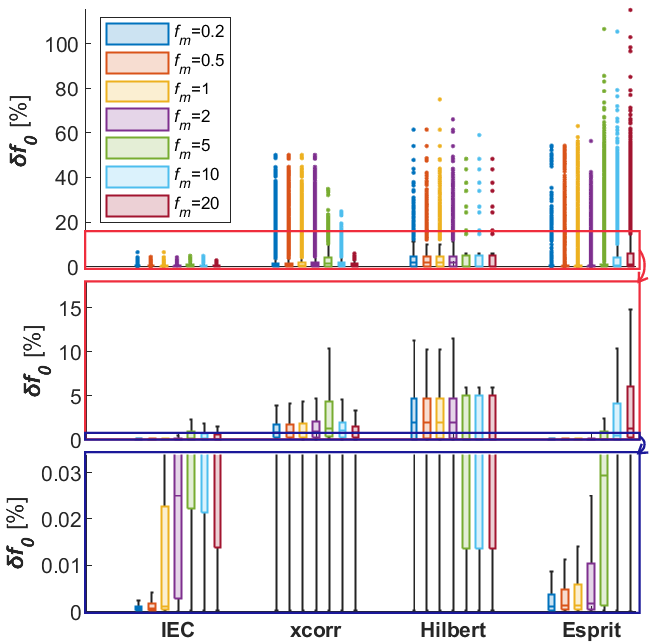}}
\caption{The dispersion of fundamental frequency estimation errors~$\delta f_0$ in the form of boxplot grouped by the modulating frequency~$f_m$ for the individual methods considered, where the top plot shows the dispersion with outliers, while the middle and bottom plots show the dispersion without the outlier with an appropriate zoom.}
\label{fig3}
\end{figure}

\begin{figure}[htbp]
\centerline{\includegraphics[width=\columnwidth]{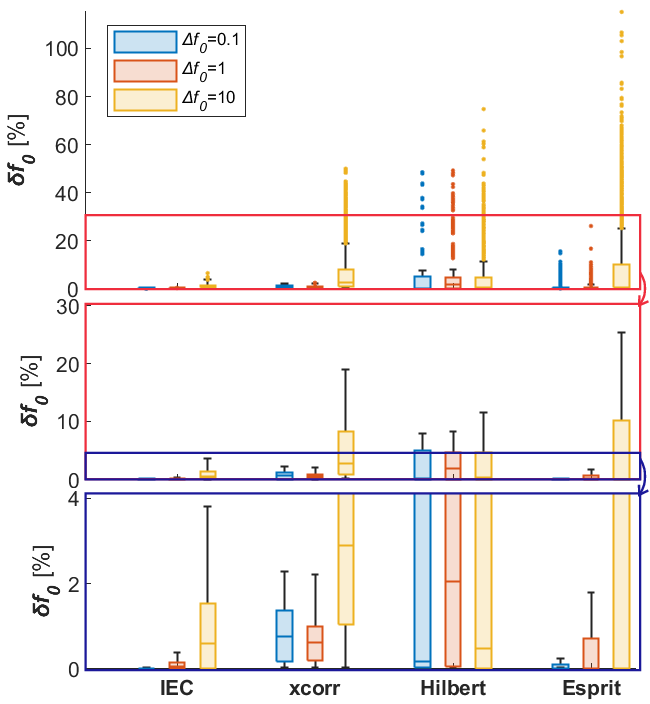}}
\caption{The dispersion of fundamental frequency estimation errors~$\delta f_0$ in the form of boxplot grouped by the deviation of fundamental frequency~$\Delta f_0$ for the individual methods considered, where the top plot shows the dispersion with outliers, while the middle and bottom plots show the dispersion without the outlier with an appropriate zoom.}
\label{fig4}
\end{figure}

\begin{figure}[htbp]
\centerline{\includegraphics[width=\columnwidth]{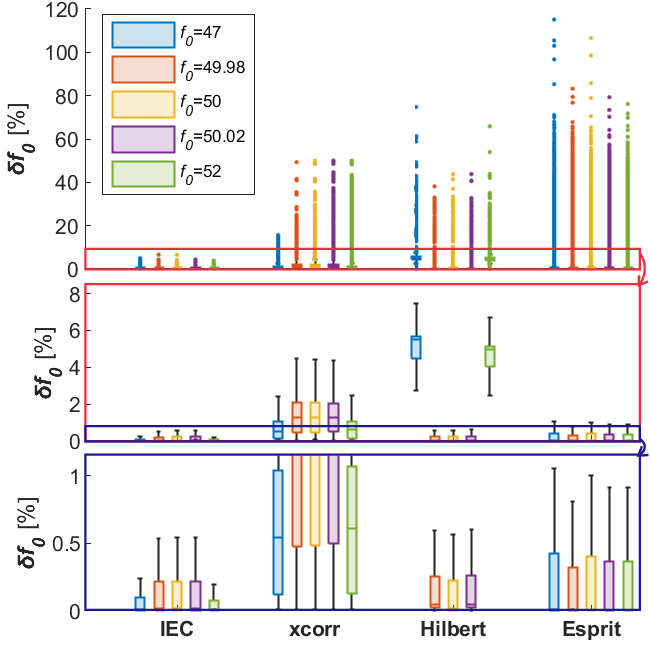}}
\caption{The dispersion of fundamental frequency estimation errors~$\delta f_0$ in the form of boxplot grouped by the fundamental frequency of the carrier signal~$f_0$ (this value is approximately equal to the average value of the fundamental frequency of the test signal determined for a 10-second window) for the individual methods considered, where the top plot shows the dispersion with outliers, while the middle and bottom plots show the dispersion without the outlier with an appropriate zoom.}
\label{fig5}
\end{figure}

\begin{figure}[htbp]
\centerline{\includegraphics[width=.95\columnwidth]{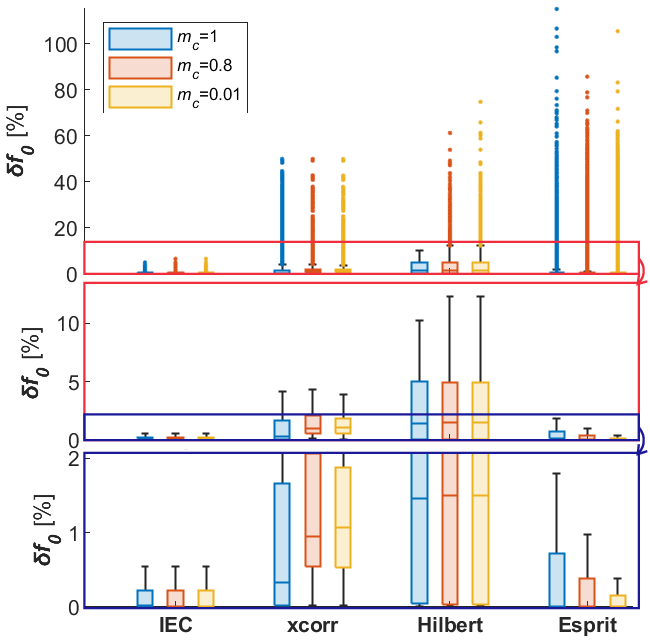}}
\caption{The dispersion of fundamental frequency estimation errors~$\delta f_0$ in the form of boxplot grouped by the clipped level~$m_c$ for the individual methods considered, where the top plot shows the dispersion with outliers, while the middle and bottom plots show the dispersion without the outlier with an appropriate zoom.}
\label{fig6}
\end{figure}

The dispersion of fundamental frequency estimation errors~$\delta f_0$ for the individual considered methods in the form of boxplot grouped by: modulating frequency~$f_m$ (rate of changes of the fundamental frequency) is shown in Fig.~\ref{fig3}, deviation of fundamental frequency~$\Delta f_0$ is shown in Fig.~\ref{fig4}, fundamental frequency of the carrier signal~$f_0$ (this value is approximately equal to the average value of the fundamental frequency of the test signal determined for a 10-second window) is shown in Fig.~\ref{fig5}, clipped level~$m_c$ is shown in Fig.~\ref{fig6} and SNR level is shown in Fig.~\ref{fig7}. Analyzing the obtained results, it can be seen that for all the methods considered, increasing the modulating frequency~$f_m$ and deviation~$\Delta f_0$ causes an increase in the fundamental frequency estimation errors. Different levels of noise and distortion have little influence on the estimation results of individual methods. Outliers of tens of percent error values can be observed for all considered test signal parameters. The best and most stable results were obtained for the IEC method, although at certain modulating frequencies, the modified ESPRIT method provided lowest error values. For the considered signal recreating simultaneous fluctuations and distortions, both the IEC method and the modified ESPRIT method provide the best results and are characterize by fundamental frequency estimation errors reduced to several percent. However, the IEC method has lower computational complexity. It is worth noting that none of the methods guarantees low method error values for the analyzed signal, which is defined in the IEC 61000-4-30 standard as $\pm$10~mHz (maximum error: ~0.02\%). Therefore, it is necessary to looking for new methods of estimating the fundamental frequency, which will guarantee obtaining results characterized by an error smaller than that defined in the relevant standard, and it is important to carry out metrological evaluation of existing measuring and recording devices by using the test signal indicated in the paper, which recreates simultaneous fluctuations and distortions.

\begin{figure}[htbp]
\centerline{\includegraphics[width=.95\columnwidth]{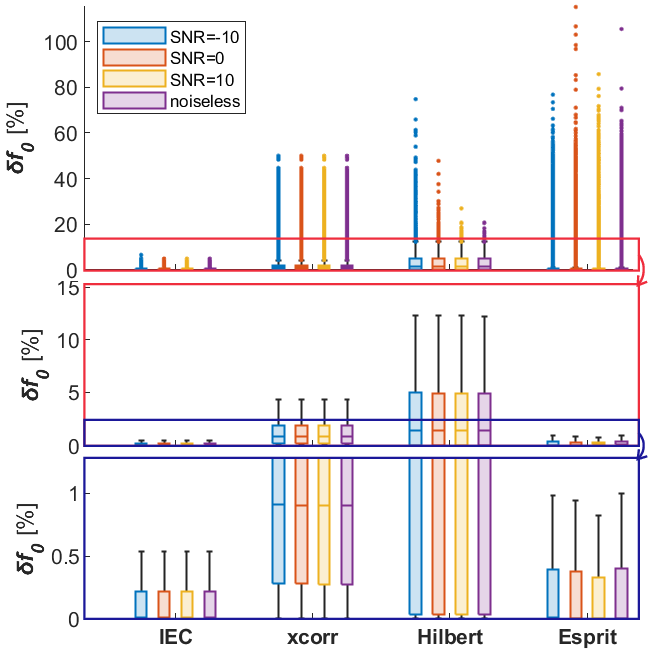}}
\caption{The dispersion of fundamental frequency estimation errors~$\delta f_0$ in the form of boxplot grouped by the SNR level for the individual methods considered, where the top plot shows the dispersion with outliers, while the middle and bottom plots show the dispersion without the outlier with an appropriate zoom.}
\label{fig7}
\end{figure}

\section{Conclusion}\label{sec5}

This paper presents the results of fundamental frequency estimation error evaluation using the currently used method specified in the IEC 61000-4-30 standard, as well as other methods available in the literature, for a test signal recreating simultaneously occurring fluctuations and distortions. The results show that none of the methods allows for error values for the analyzed signal to be lower than those required by the relevant standard. The best results are  obtained for the IEC method and the modified ESPRIT method. It is worth noting that a significant increase in estimation errors occurs with increasing frequency changes of the fundamental frequency of the measured signal and the level of variability of this frequency. The research presented in this paper indicates the need to develop a new effective method for fundamental frequency estimation and to verify the measuring and recording equipment using the test signal specified in the paper, which recreates the state of simultaneously occurring fluctuations and distortions.


\vspace{12pt}


\begin{thebibliography}{00}
\bibitem{PQRep} 6th CEER Benchmarking Report on all the Quality of Electricity and Gas Supply 2016, 2016.
\bibitem{f0_1} Verma A. K. et al., Improved Fundamental Frequency Estimator for Three-Phase Application, \textit{IEEE Transactions on Industrial Electronics}, vol. 68, no. 9, pp. 8992-8998, 2021.
\bibitem{f0_2} Li P. et al., Estimation of Non-Stationary Frequency and Fundamental Components for Power Electronics-Dominated Energy Systems, \textit{IEEE Open Journal of Power Electronics}, vol. 6, pp. 1202-1214, 2025.
\bibitem{f0_3} Reza S. et al., Accurate Estimation of Single-Phase Grid Voltage Fundamental Amplitude and Frequency by Using a Frequency Adaptive Linear Kalman Filter, \textit{IEEE Journal of Emerging and Selected Topics in Power Electronics}, vol. 4, no. 4, pp. 1226-1235, Dec. 2016.
\bibitem{AB1} Bracale A. et al., A new joint sliding-window ESPRIT and DFT scheme for waveform distortion assessment in power systems, \textit{Electric Power Systems Research}, vol. 88, pp. 112-120, 2012. 
\bibitem{AB2} Alfieri L. et al., A Wavelet-modified ESPRIT Hybrid Method for Assessment of Spectral Components from 0 to 150 kHz, \textit{Energies}, vol. 10, no. 1, art. no. 97, 2017.
\bibitem{AB3} Carpinelli G. et al., A New Advanced Method for an Accurate Assessment of Harmonic and Supraharmonic Distortion in Power System Waveforms, \textit{IEEE Access}, vol. 9, pp. 88685-88698, 2021.
\bibitem{utest} Kuwa{\l}ek P. et al., Synchronized Approach Based on Empirical Fourier Decomposition for Accurate Assessment of Harmonics and Specific Supraharmonics, IEEE Transactions on Industrial Electronics, vol. 72, no. 1, pp. 992-1002, Jan. 2025
\bibitem{Dist} Dyer S. A., Dyer J.S., Distortion: Total harmonic distortion in an asymmetrically clipped sinewave, \textit{IEEE Instrumentation \& Measurement Magazine}, vol. 14, no. 2, pp. 48-51, 2011.


\end{thebibliography}
\end{document}